\documentclass[a4paper,11pt]{article}
\usepackage{pos}
\RequirePackage{wrapfig}
\RequirePackage{subcaption}
\RequirePackage[colorinlistoftodos,prependcaption,textsize=tiny]{todonotes}

\title{Search for Stable States in Two-Body Excitations of the Hubbard Model on the Honeycomb Lattice}
\ShortTitle{Stable States in Two-Body Excitations of the Hubbard Model}

\author*[a]{Petar Sinilkov}
\author[a,b]{Evan Berkowitz}
\author[a]{Thomas Luu}
\author[c]{Marcel Rodekamp}

\affiliation[a]{Institute for Advanced Simulation (IAS-4), \\
Forschungszentrum J\"{u}lich \\
Wilhelm-Johnen-Stra{\ss}e, 52428 J\"{u}lich, Germany}

\affiliation[b]{J\"{u}lich Supercomputing Centre (JSC) and Center for Advanced Simulation and Analytics (CASA),\\
Forschungszentrum J\"{u}lich \\
Wilhelm-Johnen-Stra{\ss}e, 52428 J\"{u}lich, Germany
}

\affiliation[c]{Institut f\"{u}r Theoretische Physik, Universit\"{a}t Regensburg, \\
93040 Regensburg, Germany}

\emailAdd{p.sinilkov@fz-jelich.de}
\emailAdd{evan.berkowitz@uvi.edu}
\emailAdd{t.luu@fz-juelich.de}
\emailAdd{marcel.rodekamp@ur.de}

\abstract{We present one- and two-body measurements for the Hubbard model on the honeycomb (graphene) lattice from ab-initio quantum monte carlo simulations. Of particular interest is excitons, which are particle/hole excitations in low-dimensional systems. They are analogous to the pion in QCD, but without confinement, the question of whether they are bound and stable is of great interest in the condensed matter arena. By measuring one- and two-body correlators across various spin and isospin channels we can compute two-body energies relative to their thresholds, ultimately allowing us to check for stable states.}

\FullConference{The 41st International Symposium on Lattice Field Theory (LATTICE2024)\\
 28 July - 3 August 2024\\
Liverpool, UK\\}

\usepackage{xspace}
\usepackage{bbm}

\newcommand{\repoURL}{https://github.com/evanberkowitz/latex-base}

\newcommand{\Tabref}[1]{Table~\ref{tab:#1}\xspace}
\newcommand{\figref}[1]{Fig.~\ref{fig:#1}\xspace}
\newcommand{\Figref}[1]{Figure~\ref{fig:#1}\xspace}
\renewcommand{\eqref}[1]{(\ref{eq:#1})\xspace}

\newcommand{\goesto}{\ensuremath{\rightarrow}}

\let\builtinLaTeX\LaTeX
\def\LaTeX{\builtinLaTeX\xspace}

\begin{document}
\maketitle

\section{Introduction}\label{sec:intro}

Excitons, charge-neutral bosonic quasi-particles formed by the binding of an electron and a hole, occupy a central role in modern condensed matter physics. Their importance impacts fundamental physics as well as technological applications, including quantum computing \cite{qubits}, optoelectronics \cite{optoelectronics} \cite{organic}, and energy-efficient devices \cite{energyefficient}. Excitons are main energy carriers in materials such as perovskites, which are emerging as leading candidates for next-generation solar cells and LED technologies \cite{perovskite} \cite{Baskurt_2024}. Furthermore, excitonic phenomena have been actively studied in carbon-based bilayer systems, where tunable properties open new avenues for research and innovation \cite{bilayer} \cite{grossneveu}.

Understanding excitonic behavior poses significant theoretical challenges due to their bound-state nature, which necessitates non-perturbative approaches. Their role as analogs to pions in quantum chromodynamics (QCD), albeit without confinement, demonstrates the need for precise calculations and comprehensive analyses to unravel their electronic properties.

The Hubbard model serves as a cornerstone for investigating strongly correlated electron systems, providing a minimal yet versatile framework to study interactions in lattice structures. On a honeycomb lattice, the model captures essential features of low-dimensional systems, including Mott insulator transitions, magnetic ordering, and the emergence of Dirac cones in the electronic spectrum. These properties make it particularly suitable for exploring two-body correlations and their energy shifts relative to two-body thresholds, allowing for the systematic investigation of stable excitonic states under varying interaction strengths and system parameters.

In this work, we explore the dynamics of excitons as well as other two-body excitations in the Hubbard model on a honeycomb lattice, focusing on identifying stable states and analyzing their energy shifts. By leveraging ab initio Quantum Monte Carlo (QMC) simulations, we measure one- and two-body correlators across multiple spin-isospin channels, drawing parallels to particle interactions in QCD. This study provides insights into dynamical binding phenomena and lays the groundwork for future explorations of excitonic states in lattice systems.
\section{Honeycomb Lattice}\label{sec:hl}

\begin{figure}[htb]
    \centering
    \begin{subfigure}[b]{0.375\textwidth}
        \centering
        \includegraphics[width=100mm,scale=0.5]{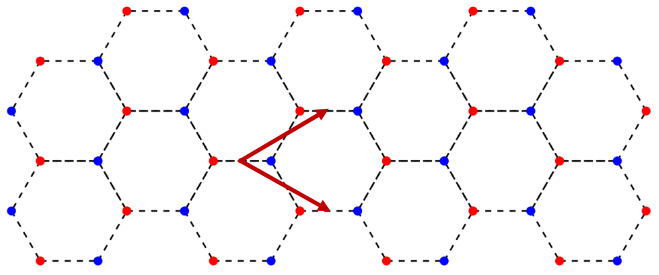}%
    \end{subfigure}
    \hfill
    \begin{subfigure}[b]{0.375\textwidth}
        \centering
        \includegraphics[width=45mm,scale=0.35]{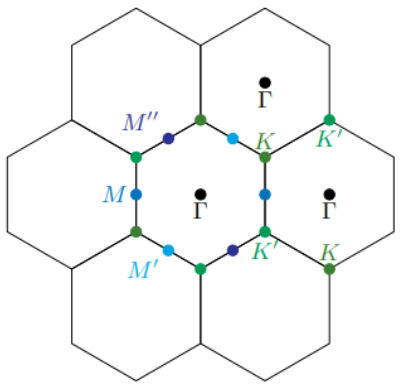}
    \end{subfigure}
    \caption{\textbf{LEFT:} Honeycomb lattice featuring color-coded sites for the two triangular sublattices. The lattice vectors are indicated by red arrows, illustrating the structural periodicity. \textbf{RIGHT:} First and second Brillouin zones of the honeycomb lattice. Key points of interest, including the center ($\Gamma$), the high-symmetry corners ($K, K'$), and the edge centers ($M, M', M''$), are highlighted for symmetry and electronic state analysis.}
    \label{fig:honeycomb}
\end{figure}

In this study, we implement the Hubbard model on a honeycomb lattice (\figref{honeycomb} left). Structurally, the honeycomb lattice consists of two triangular sublattices, making it a bipartite system, wherein each site has nearest neighbors exclusively from the opposite sublattice. This bipartite nature is foundational to the electronic and magnetic properties of the lattice, including the formation of Dirac cones at the Fermi level and the behavior under various symmetry operations.

We conduct our analysis in momentum space, restricting our focus to the first Brillouin zone (BZ), as higher zones can be mapped back into the first BZ via periodic boundary conditions. Within this zone (\figref{honeycomb} right), the key points of interest include the center ($\Gamma$), the high-symmetry corners ($K, K'$), and the edge centers ($M, M', M''$). These points serve as essential references for analyzing symmetry operations, topological features, and electronic states.

\subsection{Symmetries}

The honeycomb lattice exhibits inherent symmetries due to its periodic crystal structure. These symmetries impose constraints that must be considered in the formulation of the Hubbard model within this lattice. For instance, it is possible to leave the first BZ when adding momenta together (so-called Umklapp scattering)
\begin{equation}
    \begin{aligned}
        K + K' = \Gamma,\hspace{1cm}
        K + K = K',\hspace{1cm}
        K' + K' = K.
    \end{aligned}
\end{equation}
These relations are instrumental when applying group theory methods to classify states and interactions.

In our approach, we transition from individual momenta $k$ and $l$ of the particles to the total momentum $P$ and the relative momentum $p$ of the two-body system. This transformation leverages the conservation of total momentum
\begin{equation*}
    k, l \rightarrow P, p.
\end{equation*}
This approach allows for a more straightforward analysis of interacting particle pairs.

Additionally, we construct shells of relative momentum $p$ within the irreducible representations of the lattice's little group for a given total momentum $P$. This construction is essential for studying possible Umklapp scattering and understanding the system's interaction dynamics.

\section{Hubbard Model}\label{sec:theory}

The Hubbard model is a fundamental framework for studying interacting particles on a lattice, particularly electrons in condensed matter systems. It provides insight into phenomena such as magnetism, superconductivity, and Mott insulator transitions. Notably, the Hubbard model permits single-electron excitations, contrasting with QCD, which forbids single-quark excitations due to color confinement. This distinction highlights the unique nature of the Hubbard model's excitation dynamics. 

The Hamiltonian consists of a kinetic `tight-binding' term describing particle hopping between sites and an interaction term representing on-site electron-electron repulsion. In many cases, the Hamiltonian written in the particle-hole basis provides a more convenient form for computation. It is written as

\begin{equation}
    H = -\kappa\sum_{\langle x,y\rangle} \left(p_{x}^\dagger p_{y} - h_{x}^\dagger h_{y}\right) + \frac{U}{2}\sum_{x}q_x^2,
\end{equation}
where $p_x$ and $h_x$ represent the particle and hole operators, respectively, and $\kappa$ is the hopping parameter. The expression $\langle x,y\rangle$ denotes a nearest neighbor pair, and the sum is over all possible such pairs. The charge operator takes the form $q = n_p - n_h$, with $n_p$ and $n_h$ as the number operators for particles and holes at each site and $U$ represents the on-site interaction strength. This form simplifies certain calculations by emphasizing the duality between particles and holes.

\subsection{Non-interacting case}

For the non-interacting case ($U = 0$), an exact solution exists at half-filling ($\mu = 0$), yielding the dispersion relation \cite{physics-graphene}

\begin{equation}
	E_{\vec{k}\pm} = \pm (-\kappa)\sqrt{ 3+2\left( \cos\left(\frac{3}{2}k_x + \frac{\sqrt{3}}{2}k_y\right) + \cos\left(\frac{3}{2}k_x - \frac{\sqrt{3}}{2}k_y\right) + \cos\left(\sqrt{3}k_y\right) \right) },
    \label{eq:dispersion}
\end{equation}
where $\vec{k}$ is the momentum of the electron. This dispersion relation results in a two-band structure that can be utilized to calculate multi-particle energies. The bands meet at the Dirac points, $\vec{k}_D = \left(\frac{2\pi}{3}, \pm\frac{2\pi}{3\sqrt{3}}\right)$, forming linear energy crossings that resemble the physics of massless relativistic particles. This structure serves as a basis for studying more complex interactions.

\subsection{One-body band gap}

When $U\neq 0$, electron-electron interactions lead to correlated behavior that is absent in the non-interacting case. At large $U$, electrons experience strong on-site repulsion, which can lead to phenomena like the Mott insulator transition. Such interactions also give rise to magnetic ordering (e.g., antiferromagnetic order) at half-filling due to superexchange, a hallmark feature of the strongly interacting Hubbard model. It has been shown in the one-body states that a band gap forms \cite{pos-semimott} \cite{semimott}. It emerges at the critical coupling $U_c = 3.834(14)$. This naturally raises the question of how two-body excitations behave above and below this critical coupling, such as whether they form bound states. In this work we initially concentrate on investigating the behavior of two-body excitations slightly below the critical coupling.

\section{Correlation Functions}\label{sec:cf}
We now define the relevant interpolating operators and correlation functions which form the basis of our analysis.
We investigate two-point correlation functions in the context of the Hubbard model and quantum chromodynamics (QCD), focusing on their respective treatment of one- and two-body excitations and internal symmetry structures. Our two-point correlation functions are defined as follows
\begin{equation}
    C(\tau) = \langle \mathcal{O}(\tau) \mathcal{O}^\dagger(0) \rangle
\end{equation}
where $\mathcal{O}(\tau)$ and $\mathcal{O}^\dagger(0)$ denote time-evolved operators at time $\tau$ and time $0$, respectively.

We identify the two internal degrees of freedom within the Hubbard model as spin and isospin. The third component of isospin, $I_z$, is related to the particle's electric charge $Q = 2I_z$, where it is a matter of convention for assigning either particles or holes a positive charge.
In \Tabref{one-body}, we have shown our adopted assignment convention.

\subsection{Two-body Correlation Functions}
\begin{wraptable}{r}{5cm}
          \begin{tabular}{c| c c }
        & $S_z=\frac{1}{2}$ & $S_z=-\frac{1}{2}$ \\
        \hline
        $I_z=\frac{1}{2}$ & $p^\dagger$ & $h$\\
        \hline
        $I_z=-\frac{1}{2}$ & $h^\dagger$ & $p$\\        
    \end{tabular}
     \caption{One-body spin-isospin states}
    \label{tab:one-body}
\end{wraptable}
Building upon the one-body operators, we construct all possible two-body correlation operators, represented in \Tabref{two-body}. These operators, which are structured analogously to Clebsch-Gordan coefficients, allow us to systematically explore two-body interactions. 
In this analysis, we focus specifically on two representative interaction channels -- \eqref{repulsive} and \eqref{attractive}. We expect that channels characterized by $I_z=\pm 1$ ($Q=\pm2$) exhibit repulsive interactions, while channels with $I_z=0$ ($Q=0$) are attractive. For symmetry reasons, there is a spin degeneracy among channels within each row of the tables, as well as symmetry between positive and negative values of $I_z$.   Our simulations do indeed exhibit these symmetries.

\begin{table}[ht]
    \resizebox{.72\linewidth}{!}{
    \centering
    \begin{subfigure}[c]{0.475\textwidth}
        \caption{\textbf{$I=0, S=0$}}
        \begin{tabular}{c| c}
            & $S_z=0$ \\
            \hline
            $I_z=0$\:\:\: & $\frac{1}{2}\left( p_i p_j^\dagger + p_i^\dagger p_j + h_i h_j^\dagger + h_i^\dagger h_j \right)$
        \end{tabular}
    \end{subfigure}
        \hspace{0.2cm}   
    \begin{subfigure}[c]{0.475\textwidth}
        \caption{\textbf{$I=0, S=1$}}
        \begin{tabular}{c| c c c}
            & $S_z=1$ & $S_z=0$ & $S_z=-1$ \\
            \hline
            $I_z=0$ \:& $\frac{1}{\sqrt{2}}\left( p_i^\dagger h_j^\dagger + h_i^\dagger p_j^\dagger \right)$ & $\frac{1}{2}\left( p_i p_j^\dagger - p_i^\dagger p_j + h_i h_j^\dagger - h_i^\dagger h_j \right)$ & $\frac{1}{\sqrt{2}}\left( p_i\vphantom{^\dagger} h_j\vphantom{^\dagger} - h_i\vphantom{^\dagger} p_j \vphantom{^\dagger} \right)$
        \end{tabular}
    \end{subfigure}}
        
    \vspace{0.5cm}
    \resizebox{.72\linewidth}{!}{
        \begin{subfigure}[c]{0.475\textwidth}
            \caption{\textbf{$I=1, S=0$}}
            \begin{tabular}{c| c }
                & $S_z=0$ \\
                \hline
                $I_z=1$ & $\frac{1}{\sqrt{2}}\left( p_i^\dagger h_j - h_i p_j^\dagger \right)$ \\
                \hline
                $I_z=0$ & $\frac{1}{2}\left( p_i^\dagger p_j - p_i p_j^\dagger + h_i h_j^\dagger - h_i^\dagger h_j \right)$ \\
                \hline
                $I_z=-1$ & $\frac{1}{\sqrt{2}}\left( p_i h_j^\dagger - h_i^\dagger p_j \right)$
            \end{tabular}
        \end{subfigure}
        \hspace{0.2cm}   
        \begin{subfigure}[c]{0.475\linewidth}
            \caption{\textbf{$I=1, S=1$}}
            \begin{tabular}{c| c c c}
                & $S_z=1$ & $S_z=0$ & $S_z=-1$ \\
                \hline
                $I_z=1$ & \textcolor{cyan}{$p_i^\dagger p_j^\dagger$} & $\frac{1}{\sqrt{2}}\left( p_i^\dagger h_j + h_i p_j^\dagger \right)$ & $h_i h_j$ \\
                \hline
                $I_z=0$ & \textcolor{red}{$\frac{1}{\sqrt{2}}\left( p_i^\dagger h_j^\dagger + h_i^\dagger p_j^\dagger \right)$} & $\frac{1}{2}\left( p_i p_j^\dagger + p_i^\dagger p_j - h_i h_j^\dagger - h_i^\dagger h_j \right)$ & $\frac{1}{\sqrt{2}}\left( p_i\vphantom{^\dagger} h_j\vphantom{^\dagger} + h_i\vphantom{^\dagger} p_j\vphantom{^\dagger} \right)$ \\
                \hline
                $I_z=-1$ & $h_i^\dagger h_j^\dagger$ & $\frac{1}{\sqrt{2}}\left( p_i h_j^\dagger + h_i^\dagger p_j \right)$ & $p_i p_j$
            \end{tabular}
        \end{subfigure}
    }
    \caption{Two-body spin-isospin states.  The colored interpolating operators correspond to the ones studied in this work.}
    \label{tab:two-body}
\end{table}

We take advantage of the time-translation invariance of our system to average of multiple time-sources when performing measurements.  In some channels the $\tau=0$ (equal-time) contribution to the correlator can have delta-function contributions due to the anti-commutation relations of the fermions.  We avoid these contributions by only considering the correlation function for $\tau>0$.

Following these conventions, we can proceed with the derivation of the spin-isospin correlation functions across connected and disconnected channels. For instance:

\medskip
{\textcolor{cyan}{\underline{$I=1,S=1; I_z=1, S_z=1$:}}
\begin{equation} \label{eq:repulsive} 
{\footnotesize
    \begin{aligned}
        \left( p_i^\dagger p_j^\dagger \right) \left( p_k\vphantom{^\dagger} p_l\vphantom{^\dagger} \right) 
        &= p_i^\dagger p_j^\dagger p_k p_l + p_i^\dagger p_j^\dagger p_k p_l \\
        &= p_i^\dagger p_l p_j^\dagger p_k - p_i^\dagger p_k p_j^\dagger p_l \\
        &= p_l p_i^\dagger p_k p_j^\dagger - p_k p_i^\dagger p_l p_j^\dagger \\
        &= M^{-1}_{li}[\phi](-\tau)M^{-1}_{kj}[\phi](-\tau) - M^{-1}_{ki}[\phi](-\tau)M^{-1}_{lj}[\phi](-\tau)
    \end{aligned}
    }
\end{equation}}

{\textcolor{red}{\underline{$I=1,S=1; I_z=0, S_z=1$:}}
\begin{equation} \label{eq:attractive}  
{\footnotesize
    \begin{aligned}
        \frac{1}{2}\left( p_i^\dagger h_j^\dagger + h_i^\dagger p_j^\dagger \right)\left( h_k\vphantom{^\dagger} p_l\vphantom{^\dagger} + p_k\vphantom{^\dagger} h_l\vphantom{^\dagger} \right) \\
        = & \frac{1}{2}\left( p_i^\dagger h_j^\dagger h_k p_l + p_i^\dagger h_j^\dagger p_k h_l + h_i^\dagger p_j^\dagger h_k p_l + h_i^\dagger p_j^\dagger p_k h_l \right) \\
        = & \frac{1}{2}\left( p_l p_i^\dagger h_k h_j^\dagger - p_k p_i^\dagger h_l h_j^\dagger - h_k h_i^\dagger p_l p_j^\dagger + h_l h_i^\dagger p_k p_j^\dagger \right) \\
        = & \frac{1}{2}( M^{-1}_{li}[\phi](-\tau)M^{-1}_{kj}[-\phi](-\tau) - M^{-1}_{ki}[\phi](-\tau)M_{lj}[-\phi](-\tau) \\ 
        & - M^{-1}_{ki}[-\phi](-\tau)M^{-1}_{lj}[\phi](-\tau) + M^{-1}_{li}[-\phi](-\tau)M^{-1}_{kj}[\phi](-\tau) )
    \end{aligned}
    }
\end{equation}}
Initially, sixteen interpolating operator channels are theoretically possible: four involving disconnected diagrams (which present additional computational challenges) and twelve involving only connected diagrams. By applying symmetry considerations -- such as isospin and spin parity -- we reduce these to six independent channels, including two disconnected diagram channels.

\section{Data Analysis}\label{sec:da}

The analysis was conducted for two distinct total momenta, $\Gamma$ and $K$ (or equivalently $K'$). These specific momenta were chosen because, although $K$ is not located at the center of the Brillouin zone (BZ), its energy is lower than that of $\Gamma$, as indicated by the dispersion relation of the one-body energies \eqref{dispersion}. A side goal of this analysis was to determine whether this trend persists at the two-body level. Furthermore, we analyzed configurations where the \emph{single-body} momenta at the source and sink are denoted by $K$ and $K'$, respectively (\figref{honeycomb} right). The system under consideration is the 18-sites honeycomb lattice (with periodic boundary conditions) and interaction strength $U=3.0$. We present results across different inverse temperatures ($\beta$) and number of timeslices ($N_t$).

Due to the symmetry properties of the one-body and two-body correlation functions, a direct exponential fit to the data was not performed. Instead, we analyzed the correlators leveraging their symmetric forms:
\begin{equation}
    \begin{aligned}
        f^{1}(t) &= \sum_{n} A^1_n \cosh\left( E^{1}_{n}\left( t - \frac{\beta}{2} \right) \right),\\
        f^{2}(t) &= \sum_{n} A^2_n \cosh\left( E^{2}_{n}\left( t - \frac{\beta}{2} \right) \right) + C,
        \label{eq:fitfunc}
    \end{aligned}
\end{equation}
where $f^{1/2}(t)$ correspond to either one- or two-body correlators, $A^{1/2}_n$ represent the overlap factors, $E^{1/2}_n$ are the energies of the respective states, and $C$ accounts for the backpropagating states.
\begin{figure}[ht!]
    \centering
    \includegraphics[width=.725\textwidth]{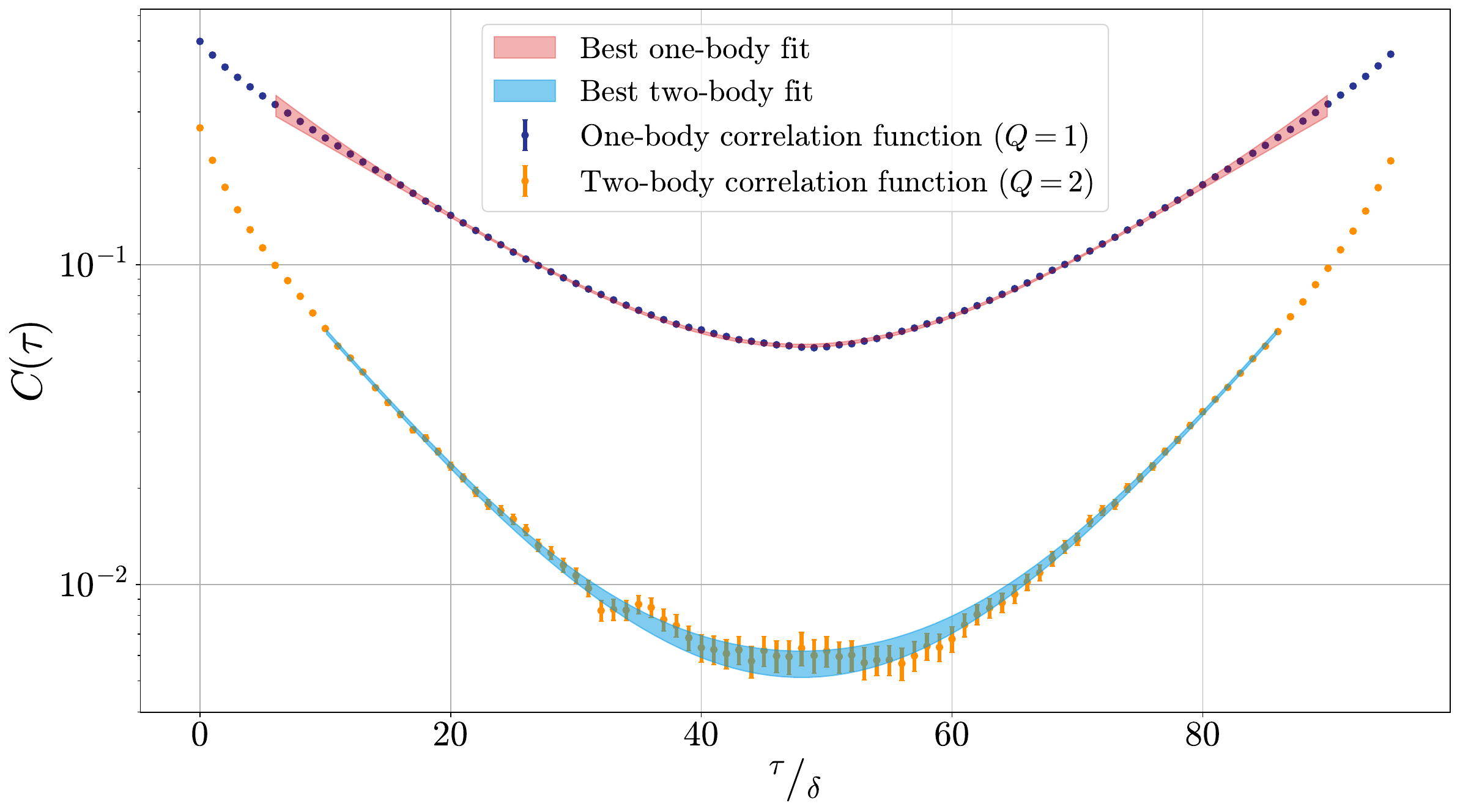}
    \caption{One- and two-body $K$-point correlators for the 18-sites honeycomb lattice with $U=3.0$, $\beta=12.0$, and $N_t=96$ with their respective best fits.}
    \label{fig:correlator_fit}
\end{figure}
We use correlated fits to these models and perform model averaving according to the AIC~\cite{perylene}. Priors are chosen from perturbative calcualtions \cite{lado}.
Using this formulation, we extract ground-state energies (\figref{correlator_fit}), which serve as a basis for calculating the energy shift, $ \Delta E_0 = E^2_0 - 2E^1_0$. The computed $\Delta E_0$ provides a quantitative measure of the energy shift in the two-body ensemble relative to the one-body threshold. Following this, the results are to be extrapolated to the continuum limit ($N_t \goesto \infty$). As these results are obtained at different $\beta$s, an extrapolation $\beta \goesto \infty$ is required to validate low-temperature behavior.
\begin{figure}[ht!]
    \centering
    \includegraphics[scale=0.212]{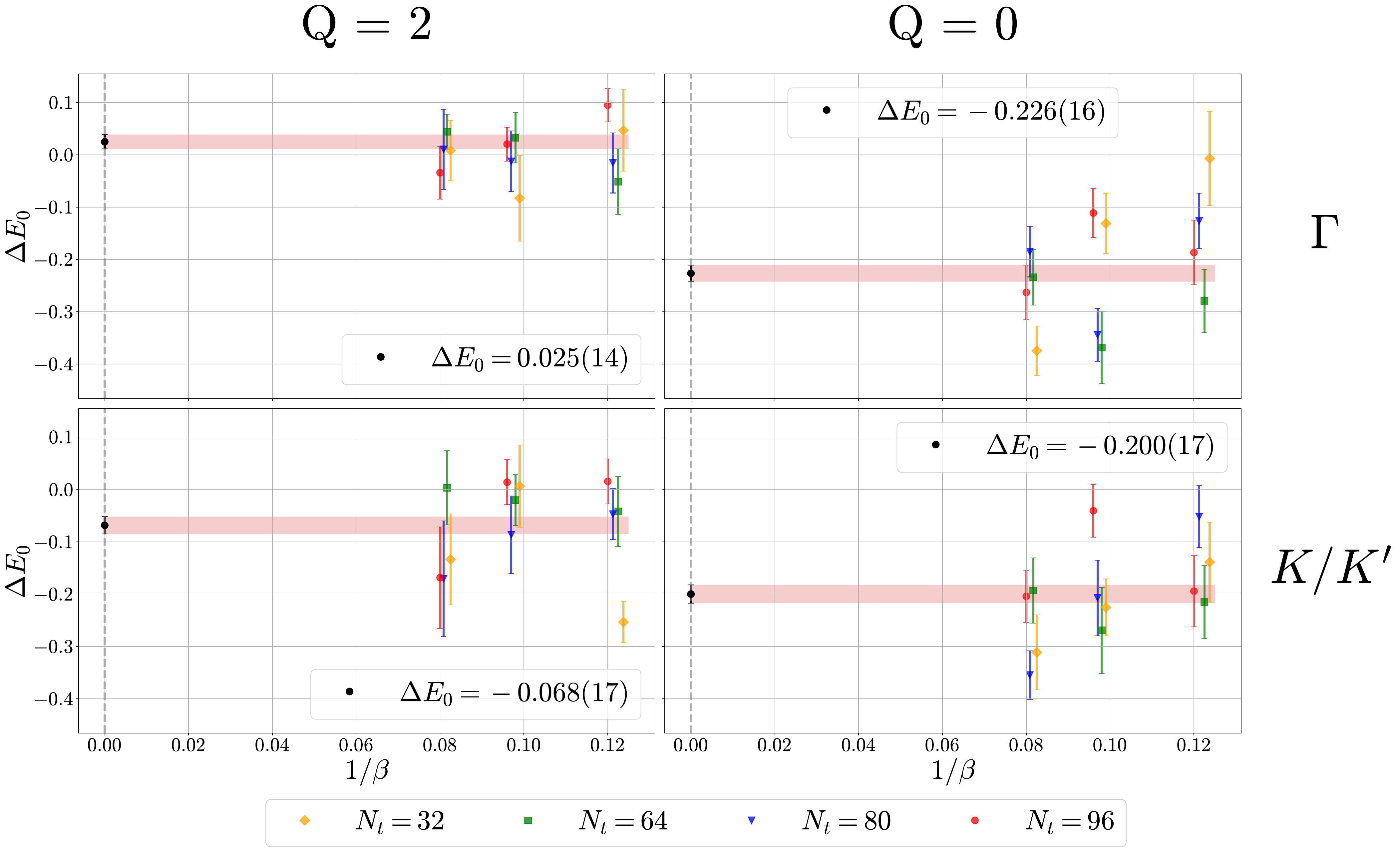}
    \caption{Energy shifts observed for two different interaction channels with distinct net charges at two total momenta and $U=3.0$. Colored markers represent data at varying timeslices, at specific inverse temperatures ($\beta$). Black points denote extrapolated results in the continuum limit at zero temperature.}
    \label{fig:energyshift}
\end{figure}

These procedures are iteratively performed for each channel and the irreducible representation under study. In this work, we focus exclusively on the $A_1$ irreducible representation, which captures the symmetric properties pertinent to the analysis. \Figref{energyshift} illustrates the energy shifts observed for two different interaction channels characterized by distinct net charges and evaluated at two separate total momenta. The colored markers in the figure represent data points obtained at various timeslices and specific inverse temperatures. Ideally, we would first perform a continuum-limit extrapolation using the results at different $N_t$, and then perform a subsequent $\beta\to\infty$ (zero-temperature) extrapolation. Unfortunately, the results for different $N_t$ are too noisy to allow for a reliable continuum extrapolation, and this in turn precludes a zero-temperature extrapolation. With the current quality of our data we can at most fit a constant to these results to estimate both continuum and zero-temperature extrapolations, as shown by the black points. We note that perturbative calculations have shown that the energy's dependence on $\beta$ is minimal, remaining nearly constant across a large range \cite{lado}, which further motivates our approach here. 

These findings highlight the consistency of the analysis framework, particularly when evaluating different interaction channels and their associated energy shifts under the symmetry constraints of the $A_1$ representation and momenta $K/K'$, $\Gamma$. This method sets a foundation for further investigation of two-body excitations and interaction effects in the Hubbard model implemented on a honeycomb lattice.
\section{Summary and Outlook}\label{sec:conclusion}

In this work, we analyzed two-body excitations in the Hubbard model implemented on a honeycomb lattice. Focusing on two representative channels, we observed that the attractive channel with net charge zero exhibited more substantial negative energy shifts compared to the repulsive channel with non-zero charge. The attractive channel demonstrated negative energy shifts across both total momenta analyzed, while the repulsive channel displayed close to zero energy shifts under the same conditions. These findings indicate that the particles within the $Q=0$ channel exhibit a stronger mutual attraction, though further investigation is needed to determine whether this shift signifies a true bound state or merely attractive interactions without binding. 

Future work will involve generating additional ensembles to reach the continuum, zero-temperature, and thermodynamic limits. Additionally, we aim to find the dependence of the energy shift on the interaction strength $\Delta E_0 \left(U\right)$ particularly for values above the critical coupling $U_c$, perform simulations at non-zero chemical potentials ($\mu \neq 0$), and improve statistical precision by incorporating more data points into extrapolations. This study provides a foundational step towards understanding excitonic behavior in lattice systems, with implications for both fundamental research and technological applications.

\acknowledgments{
  P.S. and M.R. were supported through MKW NRW under the funding code NW21-024-A. T.L was supported by the Deutsche Forschungsgemeinschaft (DFG, German Research Foundation) as part of the CRC 1639 NuMeriQS-project no.~511713970. We gratefully acknowledge the computing time granted by the JARA Vergabegremium and provided on the JARA Partition part of the supercomputer JURECA at Forschungszentrum Jülich. The authors thank Giovanni Pederiva for the valuable comments. Further, thanks go to Finn Temmen and Lado Razmadze for insightful discussions.
}

\newpage
\bibliographystyle{JHEP}
\bibliography{master}


\end{document}